\documentclass[aps,pra,amsmath,amsfonts,amssymb,floatfix,longbibliography,reprint,10pt,superscriptaddress]{revtex4-1}

\usepackage[english]{babel}
\usepackage[utf8]{inputenc}
\usepackage{graphicx}
\usepackage{tensor}
\usepackage{braket}
\usepackage{enumerate}
\usepackage{hyperref}

\usepackage{epstopdf}
\usepackage[colorinlistoftodos]{todonotes}

\newcommand{\op}[1]{\hat{#1}}                                 
\newcommand{\ketbra}[1]{\ket{#1}\!\bra{#1}}                   
\newcommand{\mean}[1]{\langle #1 \rangle}                     

\newcommand{\tr}[1]{\text{Tr}(#1)}                            
\newcommand{\ptr}[2]{\text{Tr}_{#2}(#1)}                      

\begin{document}
\title{Strengthening weak measurements of qubit out-of-time-order correlators}

\date{\today}

\author{Justin Dressel}
\affiliation{Institute for Quantum Studies, Chapman University, Orange, CA 92866, USA}
\affiliation{Schmid College of Science and Technology, Chapman University, Orange, CA 92866, USA}

\author{Jos\'e Ra\'ul Gonz\'alez Alonso}
\affiliation{Schmid College of Science and Technology, Chapman University, Orange, CA 92866, USA}

\author{Mordecai Waegell}
\affiliation{Institute for Quantum Studies, Chapman University, Orange, CA 92866, USA}

\author{Nicole Yunger Halpern}
\affiliation{Institute for Quantum Information and Matter, Caltech, Pasadena, CA 91125, USA}

\begin{abstract}
For systems of controllable qubits, we provide a method for experimentally obtaining a useful class of multitime correlators using sequential generalized measurements of arbitrary strength. Specifically, if a correlator can be expressed as an average of nested (anti)commutators of operators that square to the identity, then that correlator can be determined exactly from the average of a measurement sequence. As a relevant example, we provide quantum circuits for measuring multiqubit out-of-time-order correlators using optimized control-\emph{Z} or \emph{ZX}-90 two-qubit gates common in superconducting transmon implementations.
\end{abstract}

\maketitle

\section{Introduction}
Out-of-time-ordered correlators (OTOCs) have seen a surge of interest in recent literature due to their apparent connection to information scrambling in many-body quantum systems \cite{Larkin1969,Shenker2014,Shenker2014b,Kitaev2015,Shenker2015,Roberts2015,Roberts2015b,Hartnoll2015,Maldacena2016,Stanford2016,Maldacena2016b,Aleiner2016,Blake2016,Blake2016b,Roberts2016,Hosur2016,Lucas2016,Chen2016,Gu2017,Banerjee2017,Huang2017,Swingle2017,Fan2017,Patel2017,Chowdhury2017,He2017,Patel2017b,Kukuljan2017,Lin2018}. Prototypical systems that exhibit efficient scrambling, such as black holes, are out of reach for experimental verification, but it is still possible to simulate scrambling dynamics in the laboratory using controllable systems of qubits \cite{Swingle2016,Zhu2016,Danshita2017,Li2017,Garttner2017}. For such a simulation, an OTOC could serve as a scrambling witness. As such, there is a growing interest in measuring OTOCs for qubit systems straightforwardly.

In this paper, we extend previous work \cite{YungerHalpern2017,YungerHalpern2018} that outlines how an OTOC may be determined from a sequence of weak measurements. Such weak measurements have two shortcomings: First, they require significant data collection to overcome statistical noise. Second, they assume that backaction perturbation terms are small enough to neglect, which may be difficult to achieve experimentally. Indeed, recent experiments have found that strengthening weak measurements of other complex quantities like weak values \cite{Aharonov1988,Dressel2014} dramatically improves the accuracy of their estimation \cite{Denkmayr2017,Denkmayr2018}. To achieve similar benefits, we improve upon the sequential-measurement method by eliminating the need for weak measurements. We show how OTOCs may be exactly determined from simple averages of measurement sequences of any strength, including standard nondemolition projective measurements. 

This remarkable simplification for obtaining OTOCs with measurement sequences is restricted to observables that square to the identity, which form a useful class of observables. Many existing OTOC works consider observables with precisely this structure \cite{Hosur2016,Cotler2017,Khemani2017,Nahum2017,Nahum2018,Brown2012,Lin2017}. Such observables can have only two distinct subspaces, associated with the eigenvalues $\pm 1$, and so are natural observables to consider for practical circuit simulations using qubits. For example, the OTOC for two single-qubit observables that lie at opposite ends of a spin chain undergoing nonintegrable dynamics would be a natural short-term experimental goal \cite{Zhu2016,YungerHalpern2017,YungerHalpern2018,Swingle2018,Yao2016,GonzalezAlonso2018}.

More generally, our improved method enables the exact measurement of the expectation values of nested (anti)commutators of observables that square to the identity. Due to this generality, our method encompasses many quantities that may be of potential interest outside the field of OTOCs. We show that two-point time-ordered correlators (TOCs) and four-point OTOCs are special cases of this nested structure, and provide example circuits for how to measure these quantities. 

Since TOCs and OTOCs are complex, we use qubit measurements of two canonical types to isolate their real and imaginary parts separately: informative measurements with collapse backaction and noninformative measurements with unitary backaction. Targeting superconducting transmon qubits, we provide ancilla-based quantum circuits for implementing the two canonical qubit measurements needed to obtain the correlators. Our implementations use gates consistent with contemporary hardware and generalize experimentally prototyped methods \cite{Groen2013,Dressel2014a,White2016}.  

This paper is organized as follows. In Sec.~\ref{sec:qubitmeas} we detail the needed qubit measurement circuits and derive the general method for obtaining nested (anti)commutator averages, with supplementary details provided in the Appendix. In Sec.~\ref{sec:applications} we specialize the general result to two-point TOCs and four-point OTOCs. We summarize in Sec.~\ref{sec:conclusions}.

\section{Measuring Qubit (Anti)Commutators}\label{sec:qubitmeas}
Consider a system of controllable qubits that can be pairwise coupled with an entangling gate, assumed to be optimized for a particular hardware architecture. For concreteness, we target an array of superconducting qubits, such as transmons \cite{Koch2007,Barends2013}. Standard transmon measurements couple to the energy basis as the computational basis such that the ground state is $\ket{0}$ and the first excited state is $\ket{1}$. The qubit Pauli observables are defined as $\op{Z} = \ketbra{1} - \ketbra{0}$, $\op{Y} = -i\ket{1}\!\bra{0} + i\ket{0}\!\bra{1}$, and $\op{X} = \ket{1}\!\bra{0} + \ket{0}\!\bra{1}$, with respective eigenstates $\ket{z\pm}=\ket{1/0}$, $\ket{y\pm} = (\ket{1} \pm i \ket{0})/\sqrt{2}$, and $\ket{x\pm} = (\ket{1} \pm \ket{0})/\sqrt{2}$. As a cautionary note, this superconducting-qubit convention is opposite the quantum-computing convention for $0$ and $1$, to allow a qubit Hamiltonian to be written naturally as $\op{H}_q = E_1\ketbra{1} + E_0\ketbra{0} = \hbar\omega_q(\op{Z}/2) + \bar{E}\op{1}$, with positive qubit frequency $\omega_q = (E_1-E_0)/\hbar > 0$, and energy offset $\bar{E} = (E_1+E_0)/2$ at the mean qubit energy (and usually omitted). For simplicity, we assume that higher energy levels outside the qubit subspace may be safely neglected.

We assume that the single-qubit gates at our disposal will be the three basic rotations, $\op{R}_x(\phi) = \exp[-i(\phi/2)\op{X}]$, $\op{R}_y(\phi) = \exp[-i(\phi/2)\op{Y}]$, and $\op{R}_z(\phi) = \exp[-i(\phi/2)\op{Z}]$. These are typically implemented with optimized microwave pulses resonant with the qubit frequency \cite{Koch2007} or with a flux-bias line that tunes the qubit energy \cite{Barends2013}. We also assume that a particular two-qubit entangling gate has been optimized to match the chip geometry. We consider both the control-\emph{Z} gate \cite{Ghosh2013,Martinis2014}, $\widehat{CZ} = \ketbra{1}\otimes\op{Z} + \ketbra{0}\otimes\op{1}$, and the \emph{ZX}-90 (cross-resonance) gate \cite{Rigetti2010,Chow2011}, $\widehat{ZX}_{90} = \exp(-i(\pi/4)\op{Z}\otimes\op{X})$, as the most actively used two-qubit gates for superconducting transmon chips.

Our task is to measure multitime correlators, such as two-point TOCs $\mean{\op{B}(t)\op{A}(0)}_\rho$ or four-point OTOCs $\mean{\op{W}^\dagger(t)\op{V}^\dagger(0)\op{W}(t)\op{V}(0)}_\rho$. We will show that these correlators can be obtained exactly using temporal sequences of generalized measurements of any strength. Such a correlator generally has real and imaginary parts, which must be measured separately. To access both parts of such a correlator, we need two canonical types of measurement that probe the dual aspects of a (dimensionless) observable : 
(i) an informative measurement that causes a partial collapse onto the basis of $\op{A}$ and (ii) a noninformative measurement that causes a stochastic unitary rotation generated by $\op{A}$.  
It will become clear how these measurements enable access to real and imaginary parts, respectively, of a correlator. 

\subsection{Canonical qubit measurements}
\begin{figure*}[t]
  \includegraphics[width=2.05\columnwidth]{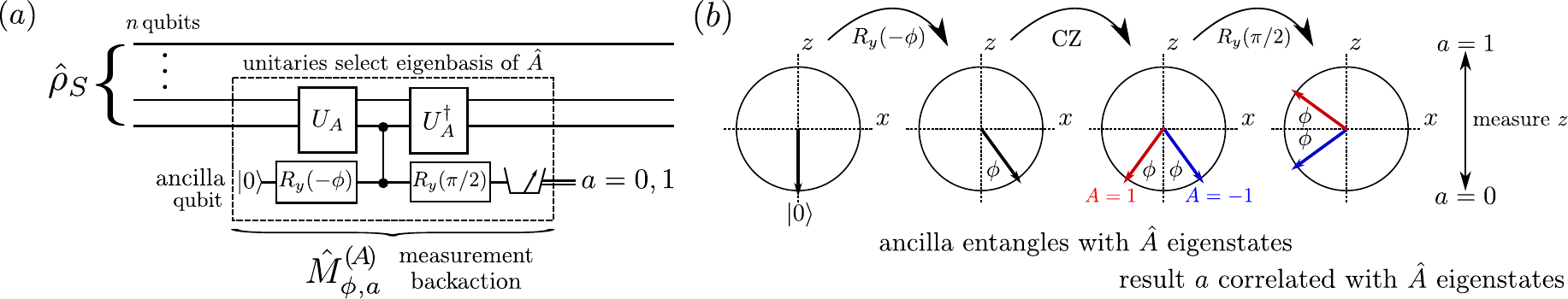}
  \caption{(a) Quantum circuit using an optimized control-$Z$ (\emph{CZ}) entangling gate to implement the generalized $\op{A}$ measurement $\op{M}^{(A)}_{\phi,a} = [\cos(\phi/2)\op{1} - (-1)^a\sin(\phi/2)\op{A}]/\sqrt{2}$. The (potentially $n$-qubit) unitary gate $\op{U}_A$ is chosen so that $\op{U}_A\op{Z}\op{U}^\dagger_A = \op{A}$ on the target qubits. The $y$-rotation gate $\op{R}_y(\varphi) = \exp[-i(\varphi/2)\op{Y}]$ rotates the ancilla qubit through an angle $\varphi$ in the $xz$-plane of the Bloch sphere. (b) Bloch-$xz$-plane detail of the ancilla evolution, showing each possible ancilla state in the entangled superposition as a distinct colored arrow. The ancilla $z$-measurement result $a=0,1$ is correlated with the eigenstates of the observable $\op{A}$, with perfect correlation when $\phi = \pi/2$. This correlation results in a partial collapse into the $\op{A}$ eigenstates. For any correlation strength $\phi$, the observable's expectation value can be determined empirically by averaging the scaled values $\alpha_{\phi,a} = (-1)^{a+1}/\sin\phi$ due to the operator identity $\sum_a \alpha_{\phi,a}\,\op{M}^{\dagger(A)}_{\phi,a}\op{M}^{(A)}_{\phi,a} = \op{A}$. }
  \label{fig:measre}
\end{figure*}

\begin{figure*}[t]
  \includegraphics[width=1.95\columnwidth]{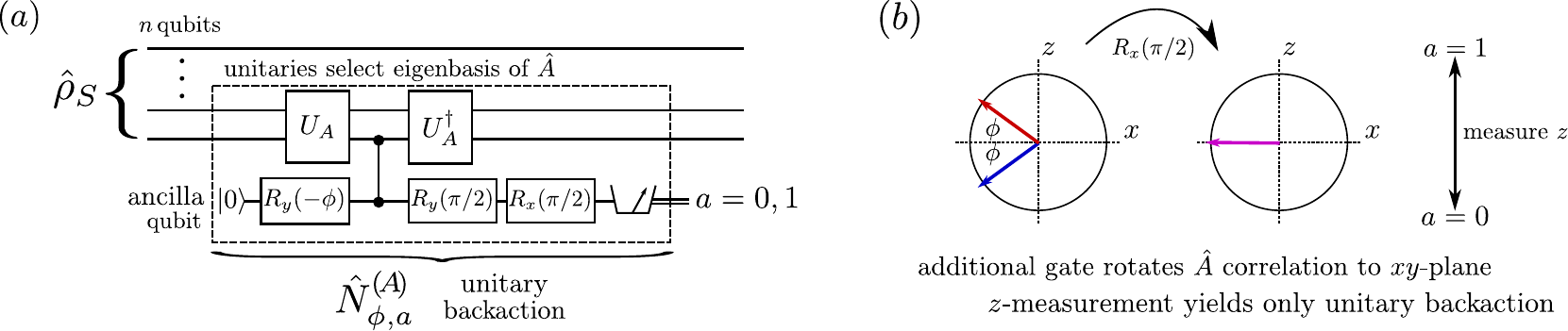}
  \caption{(a) Quantum circuit using a \emph{CZ} gate to implement the noninformative generalized $\op{A}$ measurement $\op{N}^{(A)}_{\phi,a} = [\cos(\phi/2)\op{1} + i(-1)^a\sin(\phi/2)\op{A}]/\sqrt{2}$, for comparison with Fig.~\ref{fig:measre}. The only difference is the added $x$-rotation gate $\op{R}_x(\pi/2) = \exp[-i(\pi/4)\op{X}]$ that rotates the ancilla qubit through an angle $\pi/2$ in the $yz$-plane. (b) Bloch-$xz$-plane detail of the ancilla evolution. The added rotation moves the $\op{A}$ correlation to the $xy$-plane, so the $z$-measurement result $a=0,1$ is no longer informative. Despite the lack of correlation, each result $a$ enacts a conditional unitary, generated by $\op{A}$, on the target. }
  \label{fig:measim}
\end{figure*}

\begin{figure*}[t]
  \includegraphics[width=2.05\columnwidth]{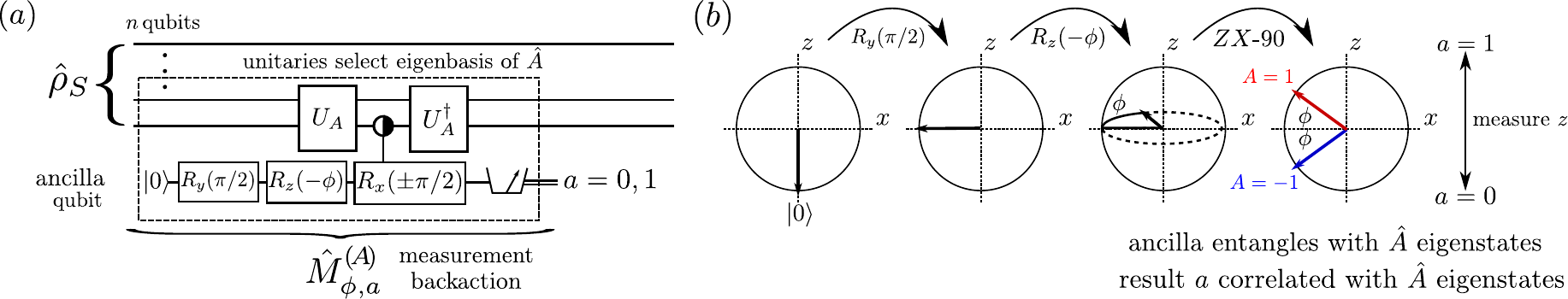}
  \caption{(a) Quantum circuit using an optimized $\widehat{ZX}_{90}$ (\emph{ZX}-90) entangling gate to implement the generalized $\op{A}$ measurement $\op{M}^{(A)}_{\phi,a}$, for contrast with Fig.~\ref{fig:measre}. (b) Bloch $xz$-plane detail of the ancilla evolution.}
  \label{fig:measrezx}
\end{figure*}

\begin{figure*}[t]
  \includegraphics[width=1.85\columnwidth]{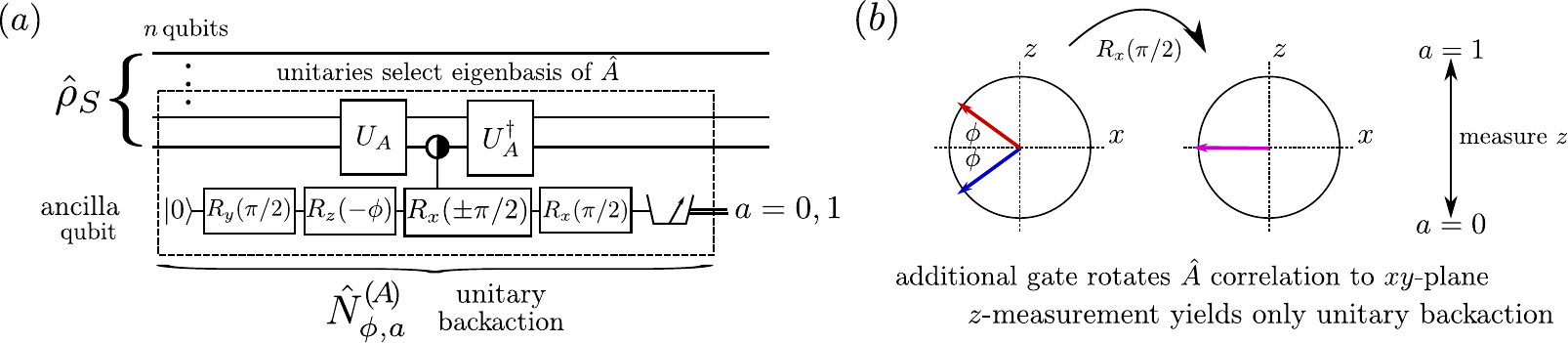}
  \caption{(a) Quantum circuit using a \emph{ZX}-90 gate to implement the noninformative generalized $\op{A}$ measurement $\op{N}^{(A)}_{\phi,a}$, for comparison with Fig.~\ref{fig:measim}. (b) Bloch $xz$-plane detail of the ancilla evolution. }
  \label{fig:measimzx}
\end{figure*}

As detailed in the Appendix, provided that an $n$-qubit operator $\op{A}$ squares to the identity $\op{A}^2 = \op{1}$ (e.g., as used in \cite{Hosur2016,Cotler2017,Khemani2017,Nahum2017,Nahum2018,Brown2012,Lin2017,Zhu2016,YungerHalpern2017,YungerHalpern2018,Swingle2018,Yao2016,GonzalezAlonso2018}), both types of $\op{A}$ measurement can be implemented using a standardized coupling to a single ancilla qubit. Such an observable has only two eigenspaces corresponding to eigenvalues of $\pm 1$ and so naturally maps onto the two eigenstates of the ancilla qubit. We provide implementation circuits using a \emph{CZ} gate in Figs.~\ref{fig:measre} and \ref{fig:measim} (see also \cite{Groen2013,Dressel2014a,White2016}) as well as implementation circuits using a \emph{ZX}-90 gate in Figs.~\ref{fig:measrezx} and \ref{fig:measimzx}. Both gate implementations yield the same entangled system-ancilla joint state prior to the ancilla collapse.

These procedures' backaction on the system can be compactly described by linear Kraus operators \cite{Nielsen2011}. Below, we derive these Kraus operators from minimal descriptions of Figs.~\ref{fig:measre}--\ref{fig:measimzx}.
\begin{enumerate}
\item \emph{Informative \textbf{M}easurement of $\op{A}$}: \\
Prepare the ancilla in the $\ket{x-}$ state, perform an $\op{A}$-controlled $y$ rotation of the ancilla through an angle $\phi$, and then measure the ancilla in the $z$ basis
  \begin{align}\op{M}^{(A)}_{\phi,\pm} &\equiv \bra{z\pm}\exp(-i(\phi/2)\op{A}\otimes\op{Y})\ket{x-} \\
	  &= \frac{\pm1}{\sqrt{2}}\left[\cos\frac{\phi}{2}\op{1} \pm \sin\frac{\phi}{2}\op{A}\right]. \nonumber
  \end{align}
  \item \emph{\textbf{N}oninformative Measurement of $\op{A}$}: \\
  Prepare the ancilla in the $\ket{x-}$ state, perform an $\op{A}$-controlled $y$ rotation of the ancilla through an angle $\phi$, and then measure the ancilla in the $y$ basis
  \begin{align}
  \op{N}^{(A)}_{\phi,\pm} &\equiv \bra{y\pm}\exp(-i(\phi/2)\op{A}\otimes\op{Y})\ket{x-} \\
  &= \frac{1}{\sqrt{2}}\left[\cos\frac{\phi}{2}\op{1} \mp i\sin\frac{\phi}{2}\op{A}\right]e^{\pm i\pi/4}. \nonumber
  \end{align}
\end{enumerate}
The initial $\ket{x-}$ state ensures that a positive measurement result correlates with the positive eigenspace of $\op{A}$ after a positive rotation angle $\phi$ in the informative case (e.g., see Fig.~\ref{fig:measre}). For clarity, we now replace the $\pm$ notation with explicit labels, e.g., $\pm 1 \to (-1)^{1+a}$ with $a\in\{0,1\}$, which will indicate the experimental outcome obtained when measuring the indicated ancilla basis. 

The informative measurement $\op{M}^{(A)}_{\phi,a}$ is a nonunitary partial projection with a coupling-strength angle $\phi\in(0,\pi/2]$ that ranges from a near-identity transformation ($\phi \approx 0$) to a full projection ($\phi = \pi/2$). That the latter is projective follows from the condition $\op{A}^2 = \op{1}$, which implies $\op{A} = \op{\Pi}_+ - \op{\Pi}_-$ and $\op{1} = \op{\Pi}_+ + \op{\Pi}_-$ for eigenprojections $\op{\Pi}_\pm$ of $\op{A}$. In contrast, the noninformative measurement $\op{N}^{(A)}_{\phi,a}$ is a measurement-controlled unitary rotation, generated by $\op{A}$, which is determined by the same $\phi\in(0,\pi/2]$, ranging from a negligible rotation $(\phi\approx 0)$ to a maximal phase difference of $\pi$ $(\phi = \pi/2)$. This noninformative case is similar to a stochastic unitary rotation. However, the experimenter knows, through the result $a$, which of the possible unitaries occurs. For example, stochastic trajectories of a superconducting qubit undergoing a sequence of noninformative measurements (also known as ``phase backaction'' \cite{Korotkov2016}) may be unitarily reversed with appropriate feedback \cite{Korotkov2006,DeLange2014}. In both the informative and the noninformative case, $\phi\in(0,\pi/2]$ conveniently parametrizes the measurement strength, allowing the tuning of the system backaction from weak ($\phi\approx 0$) to strong ($\phi=\pi/2$).

\subsection{Qubit measurement identities}
These canonical qubit measurements result in several remarkable identities, which follow from the properties in Eqs.~\eqref{eq:contextualvals}, \eqref{eq:acomm}, and \eqref{eq:comm}, derived in the Appendix. First, we define the rescaled value that the experimenter should assign each observed ancilla outcome $a\in\{0,1\}$,
\begin{align}
  \alpha_{\phi,a} \equiv \frac{(-1)^{a+1}}{\sin\phi}.
\end{align}
The values $\alpha_{\phi,a}$ act as generalized eigenvalues of the observable $\op{A}$ \cite{Dressel2010,Dressel2012b}. That is, $\op{A}$ can be decomposed into the positive-operator-valued measure for the informative measurement
\begin{align}
 \sum_{a=0,1} \alpha_{\phi,a} \op{M}^{\dagger(A)}_{\phi,a}\op{M}^{(A)}_{\phi,a} = \op{A}.
\end{align}
As a particularly important special case, when $\phi=\pi/2$, the values $\alpha_{\pi/2,a} = (-1)^{1+a}$ reduce to the eigenvalues and the measurements are projective with $\op{M}^{(A)}_{\pi/2,a} = \op{\Pi}_a$. 

Since the probability of observing an outcome $a$ is $P(a) = \text{Tr}(\op{M}^{\dagger(A)}_{\phi,a}\op{M}^{(A)}_{\phi,a}\op{\rho})$, the expectation value of $\op{A}$ may be approximated by averaging the generalized eigenvalues over $n$ trials of the experiment, $\sum_{k=1}^n \alpha_{\phi,a_k}/n \, \to_{n\to\infty}\, \sum_a \alpha_{\phi,a}P(a) = \mean{\op{A}}$. The mean-square error of this approximation is $\sum_{k=1}^n (\alpha_{\phi,a_k} - \mean{\op{A}})^2/n^2 \leq (\sum_{k=1}^n \alpha_{\phi,a_k}^2/n)/n = 1/(n\sin^2\phi)$ since $\alpha_{\phi,a}^2$ is the same for all $a$, which gives an upper bound on the root-mean-square (rms) error of $1/(\sqrt{n}|\sin\phi|)$ for the estimated mean. Strong measurements with $\phi=\pi/2$ have the smallest rms error. To guarantee the same rms error as for $n$ strong measurement trials, less strong measurements with $\phi < \pi/2$ require $n/\sin^2\phi$ trials, but also disturb the state correspondingly less. 

Typically, determining complex quantities like operator correlators requires the use of weak measurements ($\phi\approx 0$) to prevent state disturbance \cite{YungerHalpern2018,Aharonov1988}. In special cases, however, relevant information may still be contained in the collected measurement statistics in spite of any state disturbance \cite{Denkmayr2017,Denkmayr2018,Cohen2018}. In the Appendix, we show that this is the case for qubits, where the following remarkable identities hold for any coupling-strength angle $\phi$ and thus enable the improved correlator measurement protocols that are detailed in the following sections: (a) the anticommutator identities
	\begin{subequations}
\begin{align}
 \sum_{a=0,1} \alpha_{\phi,a} \op{M}^{(A)}_{\phi,a}\op{\rho}\op{M}^{\dagger(A)}_{\phi,a} &= \frac{\{\op{A},\op{\rho}\}}{2}, \\
 \sum_{a=0,1} \alpha_{\phi,a} \op{M}^{\dagger(A)}_{\phi,a}\op{B}\op{M}^{(A)}_{\phi,a} &= \frac{\{\op{B},\op{A}\}}{2} 
\end{align}
	\end{subequations}
and (b) the commutator identities
	\begin{subequations}
\begin{align}
 \sum_{a=0,1} \alpha_{\phi,a} \op{N}^{(A)}_{\phi,a}\op{\rho}\op{N}^{\dagger(A)}_{\phi,a} &= \frac{[\op{A},\op{\rho}]}{2i}, \\
 \sum_{a=0,1} \alpha_{\phi,a} \op{N}^{\dagger(A)}_{\phi,a}\op{B}\op{N}^{(A)}_{\phi,a} &= \frac{[\op{B},\op{A}]}{2i}.
\end{align}
	\end{subequations}
We show both the Schr\"odinger picture state-update forms and the Heisenberg picture operator-update forms for completeness and later convenience. For the projective case of $\phi=\pi/2$, any nondemolition projective measurement may be substituted for the ancilla measurements, making the above identities widely applicable.

These key results show that both generative aspects of an observable $\op{A}$ can be probed directly using its generalized eigenvalues: anticommutators generate nonunitary collapse backaction, while commutators generate unitary rotation backaction. We will see that the anticommutators can be used to obtain the real parts of operator correlators, while the commutators will additionally be needed to obtain the imaginary parts.

\subsection{Measurement sequence identities}
Consider a sequence of $m$ canonical system-qubit measurements implemented with the ancilla-based procedures established above. For each measurement $k=1,\ldots,m$, an ancilla $k$ will couple to an observable $\op{A}_k$, which may differ from other observables in the sequence. Depending on the basis measured on ancilla $k$, obtaining the result $a_k\in\{0,1\}$ will produce an effect $\op{K}^{(A_k)}_{\phi_k,a_k} \in \{\op{M}^{(A_k)}_{\phi_k,a_k}, \, \op{N}^{(A_k)}_{\phi_k,a_k}\}$. The probability of observing a particular sequence of results $(a_1,\ldots,a_m)$ has the form
\begin{align}
 P(a_1,\ldots,a_m) &= \\ &\tr{\op{K}^{(A_m)}_{\phi_m,a_m}\cdots\op{K}^{(A_1)}_{\phi_1,a_1}\op{\rho}\op{K}^{\dagger(A_1)}_{\phi_1,a_1}\cdots\op{K}^{\dagger(A_m)}_{\phi_m,a_m}}. \nonumber
\end{align}
That is, the measurement effects stack in a nested way.

Our main result is that,
\emph{Averaging the generalized eigenvalues, $\alpha_{\phi_k,a_k}$, for a sequence of informative (noninformative) qubit-observable measurements, $\op{M}^{(A_k)}_{\phi_k,a_k}$ ($\op{N}^{(A_k)}_{\phi_k,a_k}$), yields an expectation value of nested anticommutators (commutators) involving the measured observables.} 
That is, averaging all $\op{M}^{(A_k)}_{\phi_k,a_k}$ measurements yields
\begin{align}
  \sum_{a_1,\cdots,a_m\in\{0,1\}} & \alpha_{\phi_1,a_1}\cdots\alpha_{\phi_m,a_m}\,P(a_1,\ldots,a_m) = \\
  &\left\langle \frac{\{\cdots\{\{\op{A}_m,\op{A}_{m-1}\},\op{A}_{m-2}\}\cdots,\op{A}_1\}}{2^{m-1}}\right\rangle_\rho\nonumber,
\end{align}
while replacing the first measurement with $\op{N}^{(A_1)}_{\phi_1,\tilde{a}_1}$ yields
\begin{align}
\sum_{\tilde{a}_1,\cdots,a_m\in\{0,1\}} & \alpha_{\phi_1,\tilde{a}_1}\cdots\alpha_{\phi_m,a_m}\,P(\tilde{a}_1,\ldots,a_m) = \\
  &\left\langle \frac{[\cdots\{\{\op{A}_m,\op{A}_{m-1}\},\op{A}_{m-2}\}\cdots,\op{A}_1]}{2^{m-2}(2i)}\right\rangle_\rho\nonumber,
\end{align}
Similarly, any mixture of $\op{M}^{(A_k)}_{\phi_k,a_k}$ and $\op{N}^{(A_\ell)}_{\phi_\ell,a_\ell}$ measurements nests the appropriate anticommutators and commutators.

Remarkably, these results are exact for all measurement-strength angles $\phi_k$. This property is specific to measurements of observables satisfying $\op{A}_k^2 = \op{1}$. All decoherence terms arising from (i) the collapses due to measurement or (ii) the dephasing from random phase kicks cancel in the weighted sums. Importantly, \emph{these correlator formulas remain valid for strong measurements, wherein $\phi=\pi/2$}. Therefore, all correlators that can be written in this form are readily accessible to experiment.

The mean-square error for measurements of nested (anti)commutators $C$ like those above has an upper bound 
\begin{align}
  &\sum_{k_1,\ldots,k_m=1}^{n_1,\ldots,n_m}\frac{(\alpha_{\phi_1,{a_1}_{k_1}}\cdots\alpha_{\phi_m,{a_m}_{k_m}} - C)^2}{(n_1 \cdots  n_m)^2} \\
  &\qquad \qquad \leq \frac{1}{(n_1\cdots n_m)(\sin^2\phi_1\cdots\sin^2\phi_m)}, \nonumber
\end{align}
where $n_1,\ldots,n_m$ are the numbers of statistical trials for the measurements in the sequence. As expected, projective measurements with $\phi_k =\pi/2$ have the minimum statistical error. Compared to sequences of weak measurements with $\phi_k\approx 0$, the number of trials required for sequences of strong measurements to achieve the same rms error is greatly reduced.

\section{Applications}\label{sec:applications}
Consider measuring an operator $\op{B}(t) = \op{U}^\dagger_t\op{B}\op{U}_t$ that is evolved in the Heisenberg picture. Since $\op{B}(t)^2 = \op{U}^\dagger_t\op{B}^2\op{U}_t$, by unitarity, if $\op{B}^2 = \op{1}$, its Heisenberg-evolved version also satisfies $\op{B}(t)^2 = \op{1}$. This means all results derived in the preceding section can be applied to $\op{B}(t)$. Moreover, although the circuits in Figs.~\ref{fig:measre}--\ref{fig:measimzx} ostensibly show coupling of the ancilla to single-qubit operators, any combination of entangling unitary gates $\op{U}$ may be added before and after, to create an effective ancilla coupling to desired multiqubit operators.

Armed with these generalizations of the preceding results, we now consider two poignant examples: measuring two-point TOCs and measuring four-point OTOCs.

\subsection{Measuring two-point TOCs}
First, we consider the simple example of how to measure the two-point TOC $\mean{B(t)A}_\rho$. Suppose one starts the system in a state $\op{\rho}$, then applies a unitary evolution $\op{U}_t$, then performs a measurement $\op{M}^{(B)}_{\phi,b}$, and then applies an inverse unitary evolution $\op{U}^\dagger_t$ to obtain $\op{U}^\dagger_t\op{M}^{(B)}_{\phi,b}\op{U}_t\op{\rho}(\cdots)^\dagger$. We can group the evolutions and measurement together:
\begin{align}
\op{U}^\dagger_t\op{M}^{(B)}_{\phi,b}\op{U}_t &= \frac{\pm 1}{\sqrt{2}}\left[\cos\frac{\phi}{2}\op{1} + (-1)^{1+b}\sin\frac{\phi}{2}(\op{U}^\dagger_t\op{B}\op{U}_t)\right] \\
&= \op{M}^{(B(t))}_{\phi,b}, \nonumber
\end{align}
with a similar result for $\op{N}^{(B(t))}_{\phi,b}$. That is, performing the sequence of evolutions transforms the measurement into an effective measurement of the \emph{Heisenberg-evolved} operator $\op{B}(t)$. The linearity in $\op{B}$ of $\op{M}^{(B)}_{\phi,b}$ and $\op{N}^{(B)}_{\phi,b}$ allows for this simplification. A further simplification is obtained by noting that the cyclic property of the trace makes any final temporal evolution irrelevant for the statistical average; that is, the final inverse unitary evolution may be omitted if it is the last temporal evolution in the protocol.  

\begin{figure}[t]
  \includegraphics[width=0.7\columnwidth]{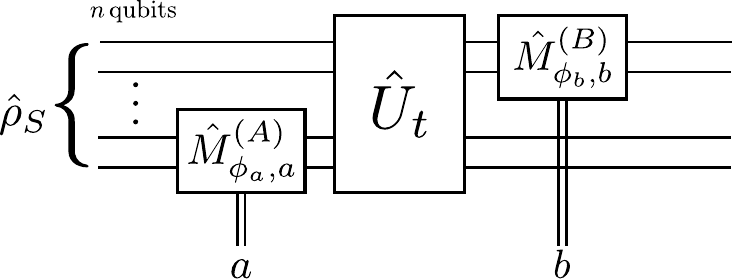}
  \caption{Quantum circuit for measuring the time-ordered correlator $\langle \op{B}(t)\op{A} \rangle_{\rho_S}$, with $\op{B}(t)=\op{U}^\dagger_t\op{B}\op{U}_t$. The operators $\op{A}$ and $\op{B}$ may act on any distinct combinations of the $n$ qubits. Using the generalized measurement procedures of any strength from Figs.~\ref{fig:measre} and \ref{fig:measrezx}, this circuit yields the distribution of results $P(a,b)$, with $a,b\in\{0,1\}$. Averaging this distribution yields $\sum_{a,b} \alpha_{\phi_a,a}\alpha_{\phi_b,b} P(a,b) = \text{Re}\langle \op{B}(t)\op{A} \rangle_{\rho_S}$, with $\alpha_{\phi_a,a}=(-1)^{1+a}/\sin\phi_a$ and similar for $b$. Replacing the first measurement with $\op{N}^{(A)}_{\phi_a,a}$ from Figs.~\ref{fig:measim} and \ref{fig:measimzx} and performing the same weighted average of results yields $\text{Im}\langle\op{B}(t)\op{A}\rangle_{\rho_S}$. }
  \label{fig:meastoc}
\end{figure}

We can therefore measure the two-time correlator with the following procedure: (i) Measure $\op{M}^{(A)}_{\phi_a,a}$. (ii) Evolve under $\op{U}_t$. (iii) Measure $\op{M}^{(B)}_{\phi_b,b}$. (iv) Average the collected distribution $P(a,b)$ of ordered result pairs $(a,b)$ with the generalized eigenvalues $\alpha_{\phi_a,a}\alpha_{\phi_b,b} = (-1)^{1+a}(-1)^{1+b}/(\sin\phi_a\sin\phi_b)$. This procedure yields the average
\begin{align}
  \sum_{a,b\in\{0,1\}}\alpha_{\phi_a,a}\alpha_{\phi_b,b}\,P(a,b) &= \left\langle \frac{\{\op{B}(t), \op{A}\}}{2}\right\rangle_\rho \\
  &= \text{Re}\langle\op{B}(t)\op{A}\rangle_\rho, \nonumber
\end{align}
which is the real part of the desired correlator. We illustrate this procedure in Fig.~\ref{fig:meastoc}. 

To find the imaginary part, only one change to the above procedure is necessary: In step (i), measure $\op{N}^{(A)}_{\phi_{\tilde{a}},\tilde{a}}$ instead, by changing the measured basis of the ancilla. Following the rest of the procedure as before yields the average
\begin{align}
  \sum_{\tilde{a},b\in\{0,1\}}\alpha_{\phi_{\tilde{a}},\tilde{a}}\alpha_{\phi_b,b}\,P(\tilde{a},b) &= \left\langle \frac{[\op{B}(t), \op{A}]}{2i}\right\rangle_\rho \\
  &= \text{Im}\langle\op{B}(t)\op{A}\rangle_\rho. \nonumber
\end{align}
Thus, both parts of the TOC may be obtained exactly using sequential measurements of any strength (including non-demolition projective measurements), without any need for reversed temporal evolution. This special case of our general qubit correlator results was also noted in Ref.~\cite{Kastner2017}.

\subsection{Measuring Pauli OTOCs}
We can use the preceding results to measure a four-point multiqubit Pauli OTOC directly in a manner similar to that of the TOC example in the preceding section. The symmetry of the OTOC expression, combined with the nice properties of the qubit Pauli operators, simplifies the nested (anti)commutators to the desired form. 

Structurally, an OTOC is the average of a group-commutator between unitary group elements $\op{V}$ and $\op{W}(t)$, where the unitary $\op{W}(t) = \op{U}^\dagger_t\op{W}\op{U}_t$ is evolved in the Heisenberg picture, like the operator $\op{B}(t)$ in the preceding TOC. Such a group commutator average has the form
\begin{align}\label{eq:otoc}
  F(t) &\equiv \mean{\op{W}^\dagger(t)\op{V}^\dagger\op{W}(t)\op{V}}_\rho
\end{align}
and measures the mean perturbations of the group operations on each other, weighted by an initial state $\op{\rho}$. Such an OTOC arises naturally from the positive Hermitian square of the algebraic commutator 
\begin{align}
  \left\langle\frac{[\op{W}(t),\op{V}]^\dagger}{(2i)^*}\frac{[\op{W}(t),\op{V}]}{2i}\right\rangle_\rho = \frac{1-\text{Re}F(t)}{2} \geq 0,
\end{align}
which implies that $\text{Re}F(t)\leq 1$.

At time $t=0$, $\op{W}(0)$ and $\op{V}$ are commonly chosen to act on independent subsystems, so that they commute and $F(0) = 1$. If, under unitary dynamics, $\text{Re}F(t)<1$, we can infer $\op{W}(t)$ has evolved to act nontrivially on the subsystem acted upon by $\op{V}$, such that $\op{W}(t)$ and $\op{V}$ do not share a common eigenbasis and thus do not commute. If the evolution is such that the $\op{W}(t)$ and $\op{V}$ nearly commute at later times, $F(t)$ will experience revivals near unity. However, nonintegrable Hamiltonian evolution can ``scramble'' local information from one subspace throughout the whole joint space such that operators on initially distinct subspaces fail to commute for very long times. Such sustained noncommutation prevents revivals in $F(t)$, making an extended absence of revivals a qualitative witness for dynamical information scrambling \cite{Larkin1969,Shenker2014,Shenker2014b,Kitaev2015,Shenker2015,Roberts2015,Roberts2015b,Hartnoll2015,Maldacena2016,Stanford2016,Maldacena2016b,Aleiner2016,Blake2016,Blake2016b,Roberts2016,Hosur2016,Lucas2016,Chen2016,Gu2017,Banerjee2017,Huang2017,Swingle2017,Fan2017,Patel2017,Chowdhury2017,He2017,Patel2017b,Kukuljan2017,Lin2018}. 

As an important special case of unitary operators for $n$-qubit systems, we will focus on separable products of Pauli operators $\op{B}(t)$ and $\op{A}$, using notation consistent with the preceding section. For example, $\op{A}$ and $\op{B}(0)$ could be local Pauli operators at opposite ends of a spin chain with nonintegrable dynamics, which is a typically considered case where an OTOC gives interesting results \cite{YungerHalpern2018}. Unitary operators of this class are Hermitian and thus satisfy $\op{A}^2 = \op{B}(t)^2 = 1$, as required to use our main qubit-measurement results. The form of the OTOC then simplifies to a four-point correlator $\mean{\op{B}(t)\op{A}\op{B}(t)\op{A}}_\rho$ similar to the preceding two-point TOC.

\begin{figure*}[t]
  \includegraphics[width=1.4\columnwidth]{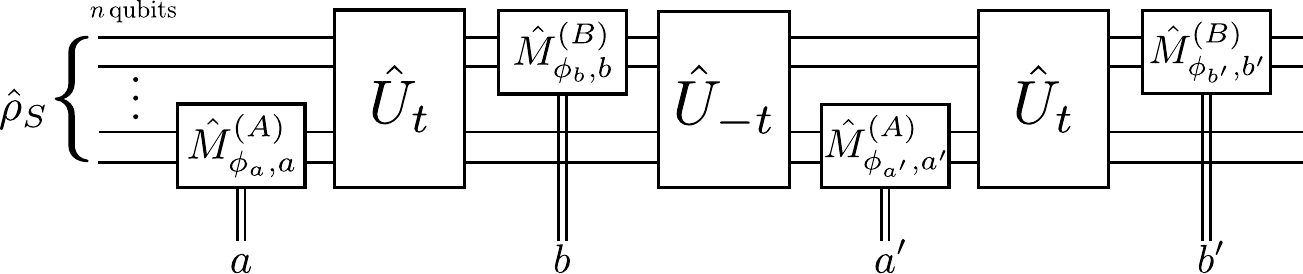}
  \caption{Quantum circuit for measuring the out-of-time-ordered correlator $F(t)=\langle \op{B}(t)\op{A}\op{B}(t)\op{A} \rangle_{\rho_S}$, with $\op{B}(t)=\op{U}^\dagger_t\op{B}\op{U}_t$. Similarly to Fig.~\ref{fig:meastoc}, this circuit yields the distribution of results $P(a,b,a',b')$, with $a,b,a',b'\in\{0,1\}$. Averaging this distribution produces $\sum_{a,b,a',b'} \alpha_{\phi_a,a}\alpha_{\phi_b,b} \alpha_{\phi_{a'},a'}\alpha_{\phi_{b'},b'} P(a,b,a',b') = (1+\text{Re} F(t))/2$, with $\alpha_{\phi_a,a}=(-1)^{1+a}/\sin\phi_a$ and similar for $b,a',b'$. Replacing the first measurement with $\op{N}^{(A)}_{\phi_a,a}$ and performing the same weighted average of results yields $\text{Im}F(t)/2$.}
  \label{fig:measotoc}
\end{figure*}

Consider the following measurement procedure: (i) Measure $\op{M}^{(A)}_{\phi_a,a}$. (ii) Evolve under $\op{U}_t$. (iii) Measure $\op{M}^{(B)}_{\phi_b,b}$. (iv) Evolve backwards under $\op{U}^\dagger_t$. (v) Measure $\op{M}^{(A)}_{\phi_a',a'}$. (vi) Evolve under $\op{U}_t$. (vii) Measure $\op{M}^{(B)}_{\phi_b',b'}$. (viii) Average the collected distribution $P(a,b,a',b')$ of ordered result quadruples $(a,b,a',b')$ with the generalized eigenvalues $\alpha_{\phi_a,a}\alpha_{\phi_b,b}\alpha_{\phi_{a'},a'}\alpha_{\phi_{b'},b'}$ [defined in Eq.~\eqref{eq:contextualvals}]. This procedure yields the average
\begin{align}
  &\sum_{a,b,a',b'\in\{0,1\}}\alpha_{\phi_a,a}\alpha_{\phi_b,b}\alpha_{\phi_{a'},a'}\alpha_{\phi_{b'},b'}\,P(a,b,a',b') \nonumber \\
  &\quad = \left\langle \frac{\{\{\{\op{B}(t), \op{A}\},\op{B}(t)\},\op{A} \}}{2^3}\right\rangle_\rho \nonumber \\
  &\quad = \frac{1 + \text{Re}\mean{\op{B}(t)\op{A}\op{B}(t)\op{A}}_\rho}{2} \nonumber \\
  &\quad = 1 - \left\langle\frac{[\op{B}(t),\op{A}]^\dagger}{(2i)^*}\frac{[\op{B}(t),\op{A}]}{2i}\right\rangle_\rho. 
\end{align}
That is, the average is precisely the complement of the Hermitian square of the commutator between $\op{A}$ and $\op{B}(t)$, which contains the real part of the desired four-point OTOC. We illustrate this procedure in Fig.~\ref{fig:measotoc}.

As with the TOC, changing only step (i) to measure $\op{N}^{(A)}_{\phi_{\tilde{a}},\tilde{a}}$ instead yields the average
\begin{align}
  &\sum_{\tilde{a},b,a',b'\in\{0,1\}}\alpha_{\phi_{\tilde{a}},\tilde{a}}\alpha_{\phi_b,b}\alpha_{\phi_{a'},a'}\alpha_{\phi_{b'},b'}\,P(\tilde{a},b,a',b') \nonumber \\
  &\quad = \left\langle \frac{[\{\{\op{B}(t), \op{A}\},\op{B}(t)\},\op{A} ]}{2^2(2i)}\right\rangle_\rho \nonumber \\
  &\quad = \frac{\text{Im}\mean{\op{B}(t)\op{A}\op{B}(t)\op{A}}_\rho}{2}, 
\end{align}
which contains the imaginary part of the same OTOC. We again emphasize that these results hold exactly for measurements of any strength.

Compared to the TOC measurement-protocol, there is a notable difference. Although we have omitted the final reverse time evolution from the protocol as before, we must perform one reverse time evolution, in step (iv). The need for this reverse evolution makes measuring the OTOC more challenging. 

Controllable qubit circuits based on gates can invert the gate sequence to reverse the evolution. If the time evolution is difficult to precisely reverse directly, a possible workaround is to introduce a time-reversal ancilla by the following extension of the Hamiltonian (inspired by the quantum-clock protocol \cite{Zhu2016}):
\begin{align}
  \op{H}_S \mapsto \op{H}_S\otimes\op{Z}.
\end{align}
If the time-reversal ancilla is in the state $\ket{1}$, time will effectively flow forward for the system as normal. If the ancilla is in the state $\ket{0}$, time will seem to flow backward for the system. This single-ancilla extension exchanges the difficulty of reversing $\op{H}_S$ with the difficulty of coupling $\op{H}_S$ to an ancilla operator $\op{Z}$.

\section{Conclusion}\label{sec:conclusions}
The sequential measurement circuits shown in this paper enable the exact determination of the expectation values of nested (anti)commutators for multiqubit observables that square to the identity. This is a useful class of observables relevant for multiqubit quantum simulations. Two-point TOCs and four-point OTOCs are special cases of this nested (anti)commutator structure, making them readily accessible to experiments with superconducting transmon qubits. Extensions to $k$-point OTOCs \cite{Roberts2017,Cotler2017,YungerHalpern2018,Haehl2017,Haehl2017b} are straightforward, but may require decomposing the $k$-point OTOC into several terms of nested (anti)commutators that could each be measured in separate experiments. Notably, measurements of any coupling strength may be used, including standard nondemolition projective measurements that minimize the statistical error. 

The method presented here improves upon the originally proposed sequential-weak-measurement approach for obtaining OTOCs \cite{YungerHalpern2017,YungerHalpern2018}. The perturbation terms now exactly cancel, avoiding the accumulated error from measurement invasiveness entirely. Moreover, using stronger measurements permits smaller statistical ensembles and less data processing. These advantages make the signal-to-noise ratio of the sequential-measurement approach now comparable to other methods to obtain an OTOC with strong measurements, e.g., the interferometric method in Ref.~\cite{Swingle2016} and the quantum-clock method in Ref.~\cite{Zhu2016}. The sensitivity of this method to experimental imperfections of the OTOC itself still requires analysis \cite{Swingle2018,Zhang2018,Yoshida2018,Yao2016,GonzalezAlonso2018}.

Although the present method is particularly useful for qubit-based simulations, the weak measurements proposed in Refs.~\cite{YungerHalpern2017,YungerHalpern2018} apply to a wider class of non-qubit OTOCs. Weak measurements also enable access to a more fundamental quasiprobability distribution (QPD) behind the OTOC \cite{YungerHalpern2018}, which we have not explored in this work. The QPD is more sensitive to measurement disturbance, and so requires more finesse to measure with arbitrary-strength measurements. 

\begin{acknowledgments}
	The authors are grateful for discussions with Brian Swingle and Felix Haehl. JD was partially supported by the Army Research Office Grant No. W911NF-15-1-0496. JRGA was supported by a fellowship from the Grand Challenges Initiative at Chapman University. MW was partially supported by the Fetzer-Franklin Fund of the John E. Fetzer Memorial Trust. NYH is grateful for funding from the Institute for Quantum Information and Matter, an NSF Physics Frontiers Center (NSF Grant No. PHY-1125565) with support of the Gordon and Betty Moore Foundation (Grant No. GBMF-2644), and for a Barbara Groce Graduate Fellowship.
\end{acknowledgments}

%

\appendix
\section{Generalized Measurement Review}\label{sec:appendix}
For completeness, we provide a full derivation of how ancilla-based measurement procedures work in a general way. We then specialize those results to qubits to show precisely where the qubit-specific simplifications arise. 

\subsection{System-ancilla coupling}
Suppose one wishes to measure a (dimensionless) observable $\op{A}$ on a \emph{system} using an ancilla \emph{detector}. One enacts a coupling gate that entangles the system's $\op{A}$-eigenbasis with the detector, and then measures the detector. The essential part of such a gate has the form
\begin{align}
\op{U}_\phi = \exp\left[-i\frac{\phi}{2}\op{A}\otimes\op{D}\right],
\end{align}
where $\phi$ is an interaction angle that dictates the coupling strength, and $\op{D}$ is a (dimensionless) detector observable. 

To see why this form creates the desired entanglement, we write the spectral expansion $\op{A} = \sum_{\lambda_A}\,\lambda_A\ketbra{\lambda_A}$ and interpret the interaction as conditionally evolving the detector state by a distinct eigenvalue-modified angle $\phi\,\lambda_A$ dependent on the eigenstate $\ket{\lambda_A}$ that the system occupies:
\begin{align}
\op{U}_\phi = \sum_{\lambda_A}\ketbra{\lambda_A}\otimes\exp\left[-i\frac{\phi\,\lambda_A}{2}\op{D}\right].
\end{align}
That is, the entangling gate is a controlled-unitary gate conditioned on the eigenbasis of $\op{A}$.

If we enact this gate on initially uncorrelated system and detector states $\op{\rho}_S\otimes\ketbra{\psi}$ and then measure a particular detector basis to obtain the result $\ket{a}$,
\begin{align}\label{eq:krausderive}
\op{\rho}_S\otimes\ketbra{\psi} &\to \op{U}_\phi\left[\op{\rho}_S\otimes\ketbra{\psi}\right]\op{U}^\dagger_\phi \\
&\to \left[\bra{a}\op{U}_\phi\ket{\psi}\op{\rho}_S\bra{\psi}\op{U}^\dagger_\phi\ket{a}\right]\otimes\ketbra{a} \nonumber \\
&\equiv \left[\op{K}^{(A)}_{\phi,a}\op{\rho}_S\op{K}^{\dagger(A)}_{\phi,a}\right]\otimes\ketbra{a}. \nonumber
\end{align}
The detector decouples from the system after the measurement yields $\ket{a}$. The resulting backaction on the system is encapsulated in the \emph{Kraus operators} \cite{Nielsen2011}
\begin{align}
\op{K}^{(A)}_{\phi,a} = \bra{a}\exp\left(-i\frac{\phi}{2}\op{A}\otimes\op{D}\right)\ket{\psi},
\end{align}
which are partial matrix elements of the joint interaction $\op{U}_\phi$. These Kraus operators effectively condition the interaction on definite detector states. For the purposes of the main text, we use notation that makes explicit the dependence of $\op{K}^{(A)}_{\phi,a}$ upon the observable $\op{A}$, the interaction angle $\phi$, and the measured detector basis $\ket{a}$, but leave implicit the dependence upon the initial detector state $\ket{\psi}$ and the coupling observable $\op{D}$, which are kept fixed in practice.

Using the spectral expansion of $\op{A}$ as before, we find
\begin{align}\label{eq:premodular}
\op{K}^{(A)}_{\phi,a} = \sum_{\lambda_A}\,\left[\bra{a}e^{-i\phi\lambda_A\op{D}/2}\ket{\psi}\right]\,\ketbra{\lambda_A},
\end{align}
so we can interpret the measurement as conditionally weighting each eigenstate of $\op{A}$ with a complex factor determined by the detector pre- and postselection $\bra{a}\,\ket{\psi}$, as well as the coupling generator $\op{D}$ and the angle $\phi$.  Factoring out the unperturbed detector amplitudes $\braket{a | \psi}$ produces the expansion $\op{K}^{(A)}_{\phi,a} = \braket{a | \psi}\sum_{\lambda_A} m^{\lambda_A}_{\phi,a} \ketbra{\lambda_A}$ in terms of the detector \emph{modular values} \cite{Kedem2010}
\begin{align}\label{eq:modularvals}
m^{\lambda_A}_{\phi,a} \equiv \frac{\bra{a}e^{-i\phi\lambda_A\op{D}/2}\ket{\psi}}{\braket{a | \psi}}
\end{align}
that completely determine how the amplitude of each $\ket{\lambda_A}$ is affected by the measurement. (If $\braket{a | \psi} = 0$, with the numerator of $m^{\lambda_A}_{\phi,a}$ nonzero for some $a$ and $\lambda_A$, $m^{\lambda_A}_{\phi,a}$ diverges, indicating that the interaction can no longer be interpreted as a multiplicative correction to the prior amplitude. One must return to the form in Eq.~\eqref{eq:premodular}.)

Generally, the detector modular values $m^{\lambda_A}_{\phi,a}$ depend upon all powers of $\op{D}$, according to the Taylor expansion of the exponential,
\begin{align}\label{eq:perturbmod}
m^{\lambda_A}_{\phi,a} = \sum_{n=0}^\infty \frac{(-i\phi\lambda_A/2)^n}{n!}\,D^{(n)}_{w,a},
\end{align}
where
\begin{align}\label{eq:weakvals}
D^{(n)}_{w,a} \equiv \frac{\bra{a}\op{D}^n\ket{\psi}}{\braket{a | \psi}}
\end{align}
are the $n$\textsuperscript{th}-order \emph{weak values} \cite{Aharonov1988} of the detector observable $\op{D}$. As we emphasized in Ref.~\cite{Dressel2014}, the perturbative series expansion in Eq.~\eqref{eq:perturbmod} is entirely specified by these weak values.

\subsection{Calibrating the measurement}
The probability of the detector result $a$ is the trace of Eq.~\eqref{eq:krausderive}: 
\begin{align}
P(a) &= \ptr{\op{K}^{\dagger(A)}_{\phi,a}\op{K}^{(A)}_{\phi,a}\op{\rho}_S}{S} \\
&= \left|\braket{a | \psi}\right|^2\sum_{\lambda_A} \left|m^{\lambda_A}_{\phi,a}\right|^2\,\bra{\lambda_A}\op{\rho}_S\ket{\lambda_A}, \nonumber
\end{align}
which implies $\mean{A} = \sum_a \alpha_a P(a)$ and the identity
\begin{align}\label{eq:contextualvals}
\op{A} = \sum_a \alpha_a \, \op{K}^{\dagger(A)}_{\phi,a}\op{K}^{(A)}_{\phi,a},
\end{align}
provided that there exist \emph{generalized eigenvalues} $\alpha_a$ that satisfy the matrix equation $\vec{\lambda} = \mathbf{C}\vec{\alpha}$, where $\vec{\lambda} = [\lambda_A]$, $\vec{\alpha} = [\alpha_a]$, and $[\mathbf{C}]_{\lambda_A,a} = |\!\braket{a | \psi}\!|^2\,|m^{\lambda_A}_{\psi,a}|^2$. A natural choice for such generalized eigenvalues is $\vec{\alpha}_0 \equiv \mathbf{C}^+\vec{\lambda}$, where $\mathbf{C}^+$ is the Moore-Penrose pseudoinverse, if it exists \cite{Dressel2010,Dressel2012b}. 

Hence, we find the general condition for being able to ``measure the system observable $\op{A}$'' in an informational sense using the ancilla detector: if Eq.~\eqref{eq:contextualvals} can be constructed by some choice of values $\alpha_a$, the detector can be calibrated to measure $\op{A}$. The generalized eigenvalues $\alpha_a$ are the values that the experimenter should assign to the empirical measurement outcomes for their statistical average to produce $\mean{A}$.

\subsection{Weak measurements}
In the case of \emph{weak coupling}, the quantity $(\phi\lambda_A)$ is sufficiently small for each $\lambda_A$ (and the $n$\textsuperscript{th}-order weak values $D^{(n)}_{w,a}$ are sufficiently well-behaved \cite{Duck1989}) to truncate this series expansion to linear order, yielding $m^{\lambda_A}_{\phi,a} = 1 - i (\phi\lambda_A/2) D_{w,a}$, where we notate $D_{w,a} \equiv D^{(1)}_{w,a}$ by convention. In this regime, the measurement's complete detector-dependence is approximately reduced to only the first-order weak value, and the Kraus operator linearizes:
\begin{align}
\op{K}^{(A)}_{\phi,a} = \braket{a | \psi}\left[\op{1} - i\frac{\phi}{2}\,D_{w,a}\,\op{A} + \mathcal{O}(\phi^2)\right].
\end{align}
It is this effective linearity in the weak regime that permits weak measurements to approximately determine multitime correlators like the OTOC, as well as quantum state amplitudes \cite{Lundeen2011} and Kirkwood-Dirac quasiprobabilities \cite{Lundeen2012,Lundeen2014} in related protocols. In particular, the change in state to order $\phi$,
\begin{align}
&\frac{\op{K}^{(A)}_{\phi,a}\op{\rho}_S\op{K}^{\dagger(A)}_{\phi,a}}{P(a)} - \op{\rho}_S \approx \\
&\; \left[\text{Re}(D_{w,a})\frac{[\op{A},\op{\rho}_S]}{2i} + \text{Im}(D_{w,a})\left(\frac{\{\op{A},\op{\rho}_S\}}{2} - \mean{A}\op{\rho}_S\right)\right]\phi, \nonumber
\end{align}
is sensitive to the commutator and/or the anticommutator of $\op{A}$ with $\op{\rho}_S$. Most importantly, relative influence can be controlled by a judicious choice of the detector weak values by manipulating the pre- and postselection states $\bra{a}\,\ket{\psi}$.

\subsection{Qubit detector and system}
In the special case of a \emph{qubit detector}, with a normalized Pauli observable $\op{D} = d_x\op{X} + d_y\op{Y} + d_z\op{Z}$ satisfying the identity $\op{D}^2 = (d_x^2 + d_y^2 + d_z^2)\op{1} = \op{1}$, the modular values in Eq.~\eqref{eq:modularvals} simplify to all orders in $\phi$,
\begin{align}
m^{\lambda_A}_{\phi,a} = \cos\frac{\phi\lambda_A}{2} - i\sin\frac{\phi\lambda_A}{2}\,D_{w,a},
\end{align}
and become completely determined by the first-order detector weak values $D_{w,a}$. The Kraus operators consequently reduce to a simpler form
\begin{align}
\op{K}^{(A)}_{\phi,a} = \braket{a | \psi}\left[\cos\frac{\phi\op{A}}{2} - i\sin\frac{\phi\op{A}}{2}\,D_{w,a}\right].
\end{align}

If the \emph{system} observable $\op{A}$ also satisfies $\op{A}^2 = \op{1}$, as for tensor products of $n$-qubit Pauli operators, the Kraus operators \emph{become linear in $\op{A}$ to all orders in} $\phi$:
\begin{align}\label{eq:qubitkraus}
\op{K}^{(A)}_{\phi,a} = \braket{a | \psi}\left[\cos\frac{\phi}{2}\op{1} - i\sin\frac{\phi}{2}\,D_{w,a}\op{A}\right].
\end{align}
This simplification allows one to achieve results similar to those in the weak-measurement regime using any coupling strength. In particular, one has the exact expression
\begin{align}\label{eq:qubitstateupdate}
\frac{\op{K}^{(A)}_{\phi,a}\op{\rho}_S\op{K}^{\dagger(A)}_{\phi,a}}{P(a)} - \op{\rho}_S &= c_{\phi,a}\,\text{Re}(D_{w,a})\frac{[\op{A},\op{\rho}_S]}{2i} \\
&+ c_{\phi,a}\,\text{Im}(D_{w,a})\left[\frac{\{\op{A},\op{\rho}_S\}}{2}-\mean{A}\op{\rho}_S\right]\nonumber \\
&+ c_{\phi,a}\frac{\sin^2\!\frac{\phi}{2}|D_{w,a}|^2}{\sin\phi}\left[\op{A}\op{\rho}_S\op{A} - \op{\rho}_S\right] \nonumber 
\end{align}
with a normalization prefactor 
\begin{align}
c_{\phi,a} = \frac{\sin\phi}{1 + \sin\phi\mean{A} \text{Im}D_{w,a} + \sin^2\!\frac{\phi}{2}(|D_{w,a}|^2-1) }
\end{align}
that generally depends on $\op{A}$. In addition to the commutator and anticommutator terms that persist in the weak regime, the third term of Eq.~\eqref{eq:qubitstateupdate} is a \emph{decoherence term} (in Lindblad form \cite{Lindblad1976}) that preserves the eigenbasis of $\op{A}$, which is the state collapse that scales with measurement strength. 

\subsection{Canonical qubit measurements}
In the main text, two strategic choices of detector configurations simplify the expressions \eqref{eq:qubitkraus} and \eqref{eq:qubitstateupdate} further. First, we set the interaction rotation to $\op{D}=\op{Y}$, to confine the detector states to the Bloch sphere's $xz$-plane. Second, we set the initial state $\ket{\psi} = \ket{x-}$ to be unbiased with respect to $z$ in that plane. Third, we choose one of two measured detector bases to select strategic detector weak values that are either imaginary or real with magnitude 1:
\begin{enumerate}
  \item $\bra{a} = \bra{z\pm} \implies D_{w,a} = \pm i$
             \[\op{K}^{(A)}_{\phi,a} \to \op{M}^{(A)}_{\phi,\pm} = \frac{\pm1}{\sqrt{2}}\left[\cos\frac{\phi}{2}\op{1} \pm \sin\frac{\phi}{2}\op{A}\right]\]
  \item $\bra{a} = \bra{y\pm} \implies D_{w,a} = \pm 1$
	  \[\op{K}^{(A)}_{\phi,a} \to \op{N}^{(A)}_{\phi,\pm} = \frac{1}{\sqrt{2}}\left[\cos\frac{\phi}{2}\op{1} \mp i\sin\frac{\phi}{2}\op{A}\right]e^{\pm i\pi/4}\]
\end{enumerate}
The overall phase factors are included for completeness but always cancel in practice.

The (unnormalized) state updates then reduce to convenient forms
\begin{widetext}
\begin{align}
\op{M}^{(A)}_{\phi,\pm}\op{\rho}_S\op{M}^{\dagger(A)}_{\phi,\pm} &= \frac{1}{2}\left[\op{\rho}_S \pm\sin\phi \frac{\{\op{A},\op{\rho}_S\}}{2} + \sin^2\!\frac{\phi}{2}\left(\op{A}\op{\rho}_S\op{A} - \op{\rho}_S\right)\right], \\
\op{N}^{(A)}_{\phi,\pm}\op{\rho}_S\op{N}^{\dagger(A)}_{\phi,\pm} &= \frac{1}{2}\left[\op{\rho}_S \pm\sin\phi \frac{[\op{A},\op{\rho}_S]}{2i} + \sin^2\!\frac{\phi}{2}\left(\op{A}\op{\rho}_S\op{A} - \op{\rho}_S\right)\right].
\end{align}
\end{widetext}
Though these expressions retain the decoherence term, it is a constant with respect to the detector outcome, while the terms of interest alternate in sign with the detector outcome. As a result, if one assigns values to the detector outcomes that also alternate in sign, then the system operations of interest can be \emph{perfectly isolated using any coupling strength $\phi$}:
\begin{align}
\label{eq:acomm}
\sum_{\pm} \left(\frac{\pm 1}{\sin\phi}\right) \op{M}^{(A)}_{\phi,\pm}\op{\rho}_S\op{M}^{\dagger(A)}_{\phi,\pm} &= \frac{\{\op{A},\op{\rho}_S\}}{2}, \\
\label{eq:comm}
\sum_{\pm} \left(\frac{\pm 1}{\sin\phi}\right) \op{N}^{(A)}_{\phi,\pm}\op{\rho}_S\op{N}^{\dagger(A)}_{\phi,\pm} &= \frac{[\op{A},\op{\rho}_S]}{2i}.
\end{align}

The operational identities in Eqs.~\eqref{eq:acomm} and \eqref{eq:comm} enable the methods in the main text. Sequential measurements nest the appropriate anticommutators and commutators, provided that all measurement outcomes are correctly averaged with alternating signs. In contrast, if early measurements in a sequence are marginalized over, the decoherence term will become important and require correction.

As a final note, Eq.~\eqref{eq:acomm} is related to the preceding notion of measuring $\op{A}$ informationally using Eq.~\eqref{eq:contextualvals}. Indeed, the average in Eq.~\eqref{eq:contextualvals} is the adjoint form of the operator update in Eq.~\eqref{eq:acomm}, provided that no subsequent measurements are performed. This relation makes it clear that the values $\alpha_{\phi,\pm} = \pm 1/\sin\phi$ in the sum are the generalized eigenvalues needed to measure $\op{A}$.

\end{document}